\documentclass[preprintnumbers,secnumarabic,amsmath,amssymb,nofootinbib,floatfix]{revtex4}
\usepackage{varioref,exscale,latexsym,amsmath,amssymb}
\usepackage{graphicx}

\usepackage{graphicx}
\usepackage{slashed}
\usepackage{dcolumn}
\usepackage{bm}
\usepackage{hyperref}
\usepackage{color}
\usepackage{xcolor}
\newcommand{\beq}{\begin{equation}}
\newcommand{\eeq}{\end{equation}}
\newcommand{\bea}{\begin{eqnarray}}
\newcommand{\eea}{\end{eqnarray}}

\def\lsi{\raise0.3ex\hbox{$<$\kern-0.75em\raise-1.1ex\hbox{$\sim$}}}
\def\gsi{\raise0.3ex\hbox{$>$\kern-0.75em\raise-1.1ex\hbox{$\sim$}}}

\def\beq{\begin{equation}}

\def\eeq{\end{equation}}

\def\beqa{\begin{eqnarray}}

\def\eeqa{\end{eqnarray}}

\begin{document}
\preprint{ACFI-T18-22}

\title{{\bf  Massive poles in Lee-Wick quantum field theory}}

\medskip\

\medskip\

\author{John F. Donoghue}
\email{donoghue@physics.umass.edu}
\affiliation{~\\
Department of Physics,
University of Massachusetts\\
Amherst, MA  01003, USA\\
 }

\author{Gabriel Menezes}
\email{gabrielmenezes@ufrrj.br}
\affiliation{~\\
Department of Physics,
University of Massachusetts\\
Amherst, MA  01003, USA\\
}

\affiliation{~ Departamento de F\'{i}sica, Universidade Federal Rural do Rio de Janeiro, 23897-000, Serop\'{e}dica, RJ, Brazil \\
 }

\begin{abstract}
Most discussions of propagators in Lee-Wick theories focus on the presence of two massive complex conjugate poles in the propagator. We show that there is in fact only one pole near the physical region, or in another representation three pole-like structures with compensating extra poles. The latter modified Lehmann representation is useful caculationally and conceptually only if one includes the resonance structure in the spectral integral.
\end{abstract}

\maketitle

\section{Introduction}

Lee and Wick have formulated a type of theory which is finite, yet yields all the usual predictions at low energy \cite{Lee:1969fy, Lee:1969fz, Lee:1970iw, Cutkosky:1969fq, Coleman, Boulware:1983vw, Grinstein:2007mp, Grinstein:2008bg}. They endow new fields with a negative metric, with the result that the propagation of these fields cancels off the high energy divergences of usual field theory. In rather simplistic terms, it is similar to including the Pauli-Villars regulators as the dynamical fields. For example the electromagnetic propagator is modified at tree level via
\beq
i D_{F\mu\nu} (q)=-i g_{\mu\nu} \left[ \frac{1}{q^2} - \frac{1}{q^2- \Lambda^2} \right] = -i g_{\mu\nu}\frac{-\Lambda^2}{q^2(q^2-\Lambda^2)} = -i g_{\mu\nu} \frac{1}{q^2 (1-\frac{q^2}{\Lambda^2})}
\eeq
The fact that the propagator goes asymptotically like $q^{-4}$ implies that loop integrals are not divergent. However, the massive field appears with negative norm - it is a ghost field. Once interactions are introduced, this dangerous feature is alleviated because the massive field decays into the light particles in the theory, such that it is not an asymptotic state in the spectrum. With some prescriptions for the treatment of loop integrals, the theory appears to be consistent and unitary, although there is a microscopic violation of causality on small scales. This Lee-Wick mechanism for dealing with theories with quartic propagators is thought to be an important ingredient for many other higher derviative theories, including that of quadratic gravity \cite{Stelle:1976gc, Julve:1978xn, Fradkin:1981hx, Tomboulis, Smilga, Einhorn, Strumia, Donoghue, Menezes, Holdom, Mannheim, Hooft, Shapiro, Narain, Anselmi, Alvarez-Gaume, Modesto, Accioly, Salvio}. It is therefore important to understand the underlying physics of Lee-Wick theories.

In these theories, when the massive state decays, the state develops a width. In most of the literature, the treatment involves a pair of poles that appear at the positions which are complex conjugates of each other, $q^2=M^2 =m_p^2+i\gamma$ and  $q^2=M^{*2} = m_p^2-i\gamma$, with $m_p^2,~ \gamma$ both real. However, when explicit calculations are needed, one finds that there is only one pole. A representation with three poles - with two of them compensating - is also valid and useful. To our knowledge, the compensation of the latter poles was first noted in the context of an $O(N)$ model by Grinstein, O'Connell and Wise \cite{Grinstein:2008bg}. The purpose of this paper is to provide a clear discussion of this issue and to highlight the importance of the spectral integral in the latter representation.

\section{Location of the pole in the propagator}

Both the usual photon and the heavy Lee-Wick particle couple to the electromagnetic current, and so the interaction can be described by a combined propagator. We start with a simple representation of this taken from the literature, which is quite intuitive and which captures the essence of the theory. We will later return to explain why this simple result is representative of the more complicated treatments that one finds when perusing the original literature. In Ref. \cite{Boulware:1983vw}, Boulware and Gross give the following representation for the propagator
\beq
i D_{F\mu\nu}(q^2)= -i g_{\mu\nu} D(q^2)
\eeq
with
\beq \label{basicform}
D(q^2) =\frac{1}{(q^2+i\epsilon)\left[ 1 + \hat{\Pi}(q^2) - \frac{q^2}{\Lambda^2}\right]}
\eeq
with
\beq
\hat{\Pi}(q^2) = q^2 \frac{\alpha}{3\pi}\int_{4m_f^2}^{\infty} ds \frac{1}{s(s-q^2-i\epsilon)} \sqrt{1-\frac{4m_f^2}{s}}\left(1+\frac{2m_f^2}{s}\right)
\eeq
Here $\hat{\Pi}(q^2)$ is the finite part of the vacuum polarization function for a fermion of mass $m_f$, written in a dispersion theory representation\footnote{See for example Problem V-3 in Ref. \cite{DSM}.}. The propagator has been modified by integrating out the fermions, and the vacuum polarization yields a well-known correction factor.  In transcribing this equation from Ref. \cite{Boulware:1983vw}, we have modified it to conform to our $(+,-,-,-)$ metric and also added the $i\epsilon$ in the denominator. This $i\epsilon$ is relevant in what follows because it helps determine the location of the pole. The location and sign of the $i \epsilon$ in the vacuum polarization is unambiguously determined by the usual placement of $i\epsilon$ in the fermion propagator.

The result is important in Lee-Wick theories because the vacuum polarization develops an imaginary part for large time-like $q^2$, which leads to a width for the massive ghost-like resonance. The fact that the massive ghost is unstable is crucial for the interpretation of the ghost, as it implies that the ghost does not appear in the asymptotic spectrum. Moreover, the other unusual properties of Lee-Wick ghosts, such as the opposite sign for the width, also follow from the form of the vacuum polarization.

\subsection{Numerical study}

First let us display the content of the propagator using the vacuum polarization without any approximation and at a moderate coupling. Using a larger coupling allows one to visibly see the properties of the propagator which are harder to display if one uses a narrow width approximation. However in this case, the mass and residue of the high mass pole must be found numerically. We will return to a useful analytic approximation at weaker coupling below.

The vacuum polarization function for a fermion of mass $m_f$ has the form
\beq
\hat{\Pi}(q^2) = -\frac{\alpha}{3\pi} \left( 1 +\frac{2 m_f^2}{q^2}\right)\left[\sigma \log \left(\frac{1+\sigma}{1-\sigma}\right) - i\pi \sigma - \frac53 \right]
\eeq
for $q^2> 4 m_f^2$, where
\beq
\sigma = \sqrt{1 -\frac{4 m_f^2}{q^2} }\ \ .
\eeq
For smaller timelike values of $0\le q^2 \le 4m_f^2$, it has the form
\beq
\hat{\Pi}(q^2) = -\frac{2\alpha}{3\pi} \left( 1 +\frac{2 m_f^2}{q^2}\right)\left[\left(\sqrt{\frac{4 m_f^2}{q^2}-1}\times {\rm arccot} ( \sqrt{\frac{4 m_f^2}{q^2}-1})-1\right)  + \frac16 \right]  \ \ .
\eeq

Let us explore the propagator using $\alpha = 0.25$, which is large enough to see the essential features. The value of $\Lambda$ can be chosen to set the overall scale of units, and our exploration sets $\Lambda = 1$. For convenience we choose the fermion mass to be $m_f^2 =0.004$ in these units, small enough that threshold effects are not important near the pole but large enough that the logarithms of $q^2/m_f^2$ do not become so large as to affect the numerical accuracy.

\begin{figure}[htb]
\begin{center}
\includegraphics[height=60mm,width=65mm]{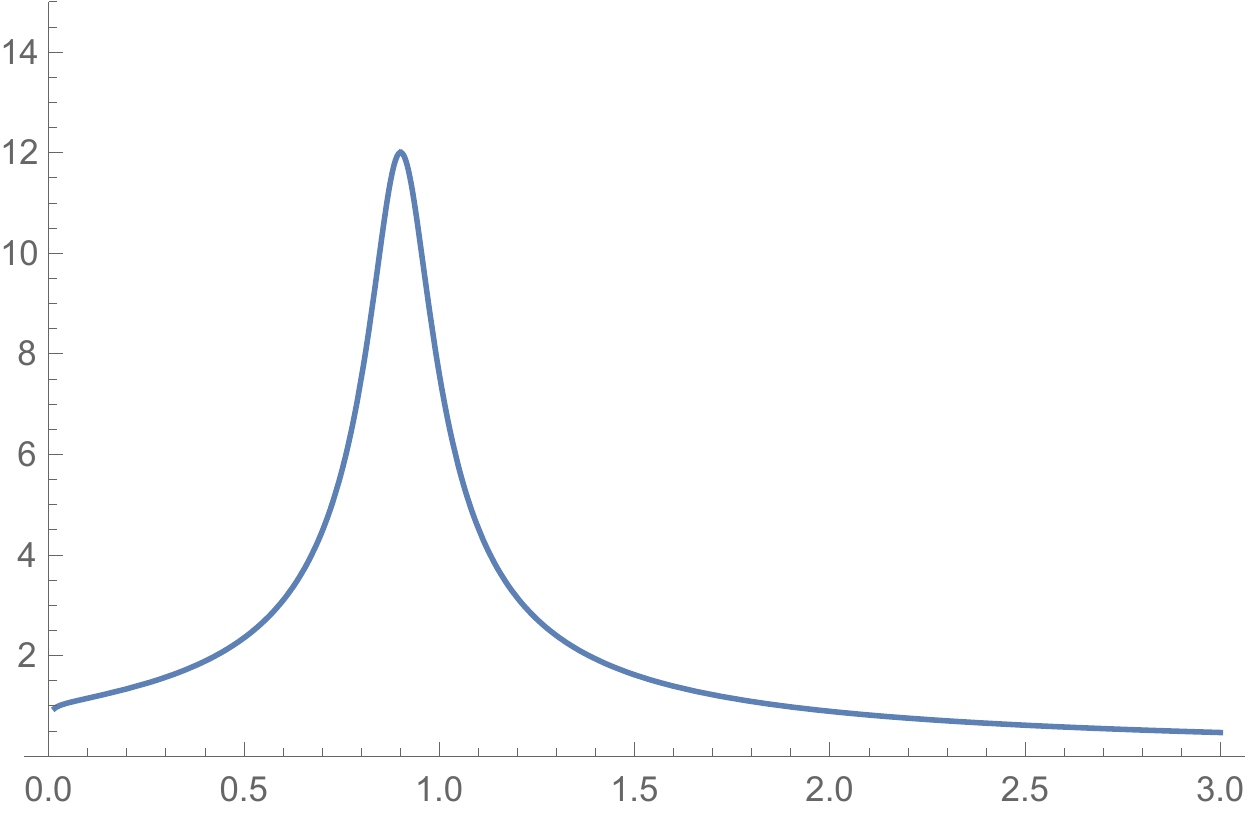}
\caption{The absolute value of the propagator as a function of $q^2$, using the parameters discussed in the text, in $\Lambda=1 $ units, multiplied by $q^2$}
\label{LWprop}
\end{center}
\end{figure}

\begin{figure}
\centering
\begin{tabular}{cc}
\includegraphics[height=30mm,width=60mm]{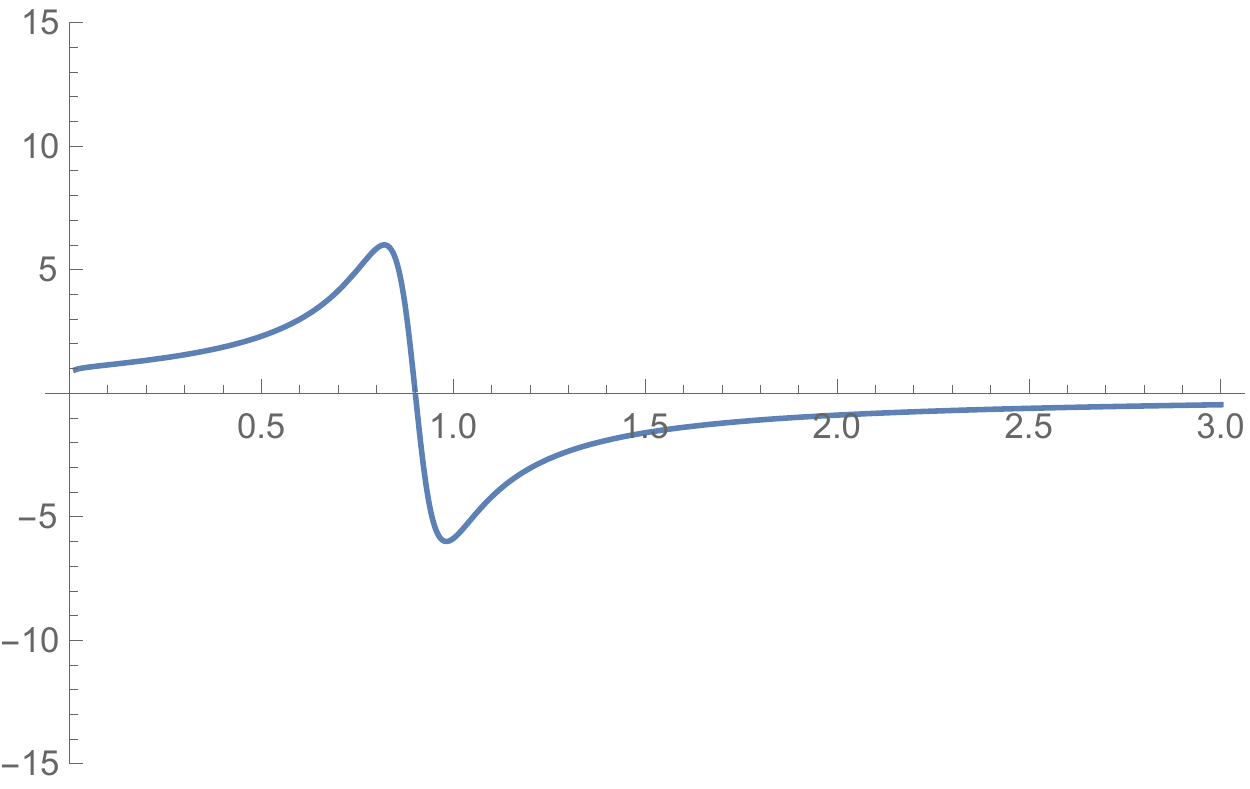}&
\includegraphics[height=30mm,width=60mm]{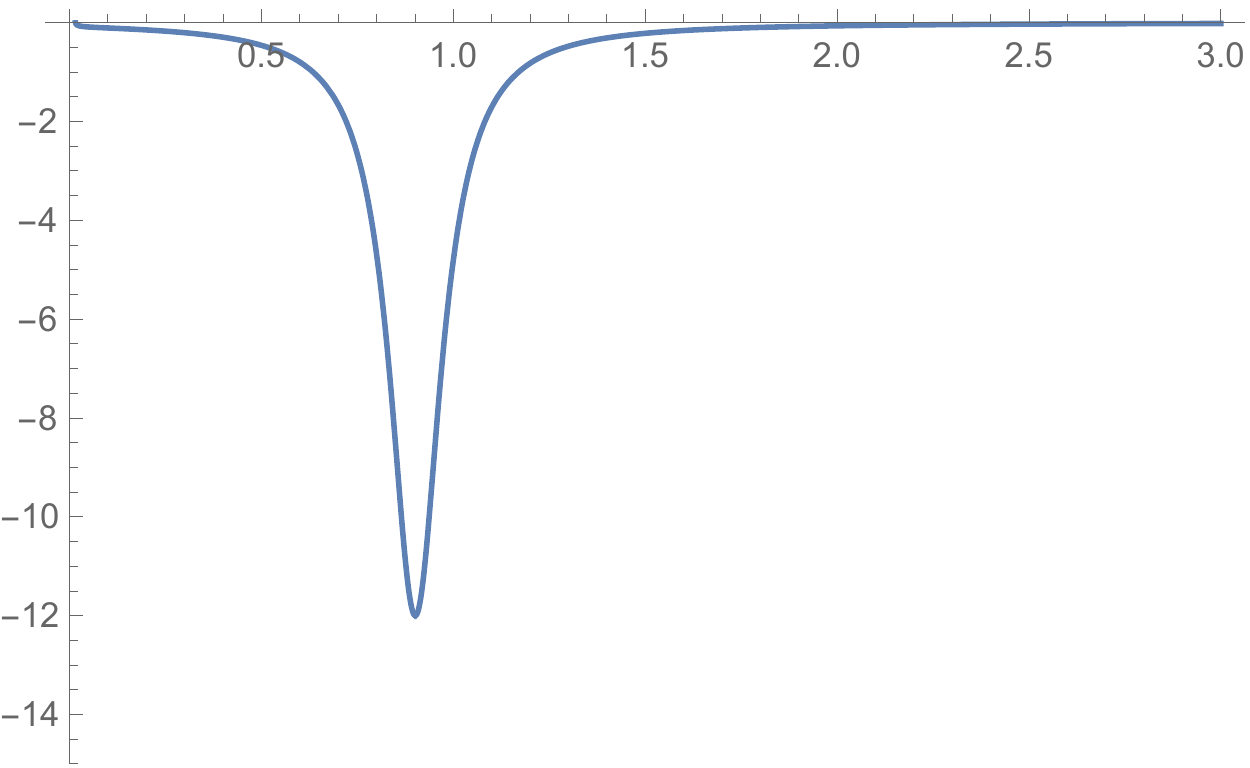}
\end{tabular}
\caption{Real and imaginary parts of the propagator, in the same format as Fig. \ref{LWprop}. }
\label{LWrealim}
\end{figure}

The absolute value of the propagator is displayed in Fig. \ref{LWprop}. Here we have multiplied by a factor of $q^2$ so that a normal photon propagator would be displayed as a flat line of magnitude unity. One can see the fall-off at high values of $q^2$ as the Lee-Wick propagator asymptotically goes as $1/q^4$. However, the most striking feature is obviously the resonance structure which is the unstable ghost resonance. The real and imaginary parts of the propagator are shown separately in Fig. \ref{LWrealim}. The zero in the real part and the peak in the imaginary part occur at $q^2=0.90116$ in $\Lambda=1 $ units. The half-width at half-maximum of the imaginary part is $0.0809$ in these units.

In the physical region, keeping $q^2$ real, we find that the resonance shape can be well approximated by the form
\beq
D(q^2) \sim \frac{-\tilde{\beta}}{q^2 - \mu^2 - i \tilde{\gamma}}
\eeq
with the parameters
\begin{eqnarray}
\mu^2 &=& 0.90116   \nonumber \\
\tilde{\gamma} &=& 0.0809 \nonumber \\
\tilde{\beta} &=& 1.0772  \ \ .
\end{eqnarray}
This propagator has two unusual signs in this region. The overall minus sign is the signal of the ghost-like nature. But also the imaginary part in the denominator is different from the usual structure for a propagator of an unstable particle. In the standard case, we normally have
\beq
D(q) = \frac{1}{q^2 - (M-i \frac{\Gamma}{2})^2} = \frac{1}{q^2 - \mu^2 +i M\Gamma }
\eeq
where in this case $\mu^2 =M^2-\Gamma^2/4$, This sign difference in the width is a well-known feature of the Lee-Wick ghost.
However we note that because of the two sign differences, the imaginary part of the overall propagator is in agreement in the two cases,
\beq
{\rm Im} D(q) =  - \frac{ \tilde{\beta} \tilde{\gamma}}{ (q^2-\mu^2)^2 +\tilde{\gamma}^2}  ~~~ vs ~~~ - \frac{M\Gamma}{(q^2-\mu^2)^2 +M^2\Gamma^2} \ \ .
\eeq
We reiterate that the sign of the width follows directly from the usual $i\epsilon$ prescription for the fermion propagator.

The parameters above were determined while keeping $q^2$ real. An alternate procedure would be to let $q^2$ become a complex variable, and to search for a pole in the propagator in the complex plane. If we follow this procedure and look in the neighborhood of the resonance of the previous paragraph, we find a pole described by the parameters
\beq
D(q^2) \sim \frac{-{\beta}}{q^2 - m_p^2 - i {\gamma}}
\eeq
with
\begin{eqnarray}\label{complex}
m_p^2 &=& 0.901045   \nonumber \\
{\gamma} &=& 0.08089 \nonumber \\
{\beta} &=& 1.06922 - 0.09308 i \ \ .
\end{eqnarray}
The slight differences in the pole position and the residue between this determination and the previous one arises because the propagator is not purely quadratic in the momentum in this region, with the presence of the logarithm being important. We note that in the modified Lehmann representation, which we will discuss next, it is important to use this latter description of the pole parameters.

\subsection{Modified Lehmann representation}

It might be tempting to represent the propagator as the sum of two poles, for example
\beq
D(q) \sim \frac{1}{q^2} - \frac{1}{q^2-\mu^2 -i\gamma}  \ \ .
\eeq
This is accurate to a certain level, but has fundamental flaws. For example, it would give the {\em spacelike} propagator an imaginary part, which violates general principles of QFT. There is however an exact representation of the propagator in terms of poles and cuts which is very useful in elucidating the physical content of the theory. To our knowledge, this modified Lehmann representation \cite{Kallen:1952zz, Lehmann:1954xi} was first introduced by Coleman in Ref. \cite{Coleman}. It has some unusual features which we aim to explain in this subsection, again using the numerical example given above.

\begin{figure}[htb]
\begin{center}
\includegraphics[height=90mm,width=90mm]{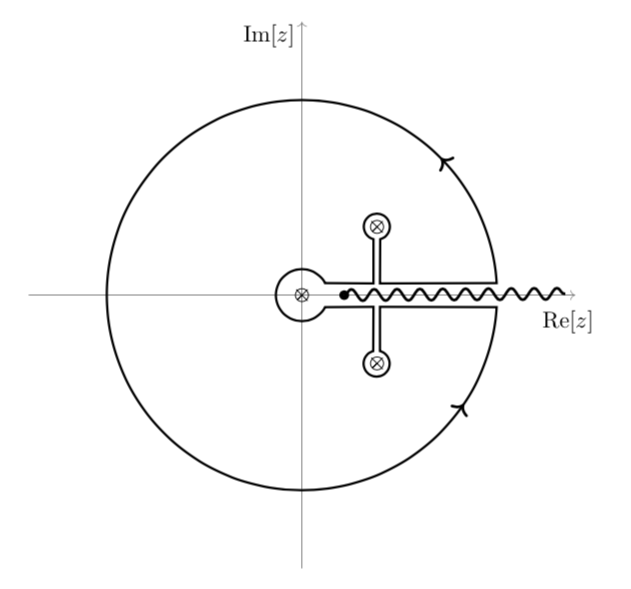}
\caption{The contour in the complex plane that yields the Lehmann representation. The poles are represented by encircled crosses, the branch cut is represented by a wavy line and the branch point is the black dot on the real axis. As discussed in the text, the cut begins at $q^2= 4m_f^2$. }
\label{figcontour2}
\end{center}
\end{figure}

If we consider the propagator as a more general function of complex $q^2$ we see four key features. There is of course the pole at $q^2=0$ and the cut running along the real axis from $q^2= 4m_f^2$ to $\infty$. We have discussed above the massive pole found slightly above the real axis. If we look on the other side of the cut, starting from $q^2-i\epsilon$ instead of $q^2+i\epsilon$, we find the complex conjugate location of the massive pole, i.e. at $q^2= m_p^2 -i\gamma$. Because the propagator falls off fast enough at $|q^2|\to \infty$, we can use the Cauchy formula
\begin{equation}
f(q^2) = \frac{1}{2\pi i} \oint \frac{f(z)}{z-q^2}dz
\end{equation}
with the contour of Figure \ref{figcontour2} in order to write the identity
\begin{equation}\label{Lehmann}
D(q) = \frac{1}{q^2+i\epsilon} - \frac{\beta}{q^2-M^2} - \frac{\beta^*}{q^2-M^{*2}} + \frac{1}{\pi}\int_{4m_f^2}^\infty ds \frac{\rho(s)}{q^2-s+i\epsilon}
\end{equation}
where $\beta, ~\beta^*$ are the residues at the massive pole and the the spectral function $\rho(s)$ is given by the discontinuity across the cut.

This representation satisfies the basic requirements of field theory. There is no imaginary part for spacelike momenta. The contribution from the massive pole is cancelled by the complex conjugate pole. The only imaginary parts arise from the spectral function integral, which has an imaginary component only for $q^2>4m_f^2$. The existence of two complex-conjugated poles is a well-known property of Lee-Wick theories. The spectral function contribution is however equally important and is much less discussed in the literature. We will see that it is also pole-like in an important way.

\begin{figure}[htb]
\begin{center}
\includegraphics[height=60mm,width=120mm]{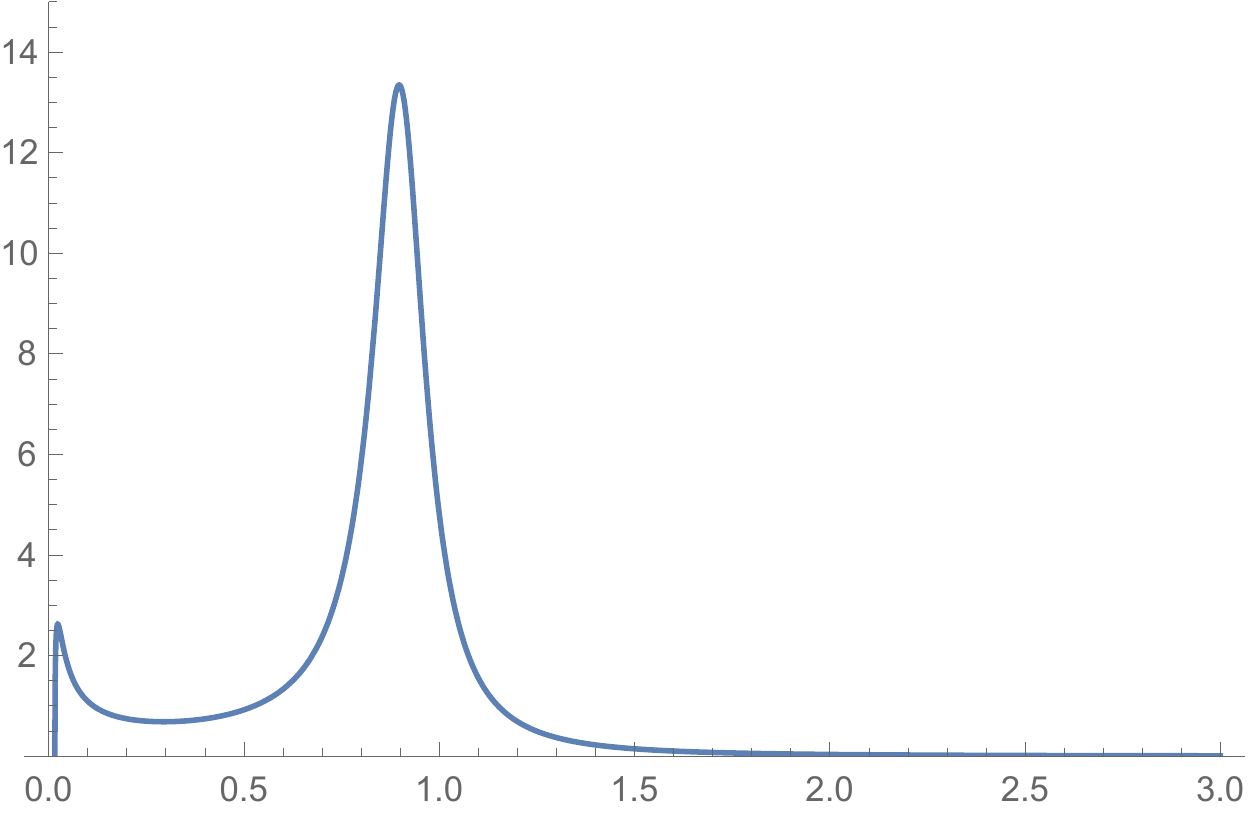}
\caption{The spectral function for $\alpha=0.25$ and $\Lambda = 1$. }
\label{spectral}
\end{center}
\end{figure}

We can isolate the spectral function by taking the imaginary part of the propagator
\beq
-{\rm Im}~  D(q) |_{q^2=s} = \rho (s)   \ \ .
\eeq
This produces
\beq
\rho(s) = \frac{{\rm Im}\hat{\Pi}(s)}{s\left[\left(1 - \frac{s}{\Lambda^2} +{\rm Re} \hat{\Pi}(s) \right)^2+ \left({\rm Im} \hat{\Pi}(s)\right)^2\right]}
\eeq
In the region near the massive pole, the imaginary part of the vacuum polarization is approximately a constant
\beq
{\rm Im} \hat{\Pi}(s) \sim -\frac{\alpha}{3}   ~~~~~~~~~ {\rm for} ~~q^2>> 4 m_f^2   \ \ .
\eeq
This gives the spectral function a shape which is very close to a Breit-Wigner form. The result is shown in Fig. \ref{spectral} for the parameters of the previous subsection. We have verified numerically that the use of this spectral function along with the pole results given by the determination of Eq. \ref{complex} reproduces the real and imaginary parts of the propagator within the numerical accuracy of the calculation.

These results appear to raise a puzzle. Our exploration of the original propagator revealed a single pole which governed the behavior in the physical region. The modified Lehmann representation involves three massive pole-like structures, i.e. the pole and the complex conjugate pole as well as the spectral function. How can these describe the same physical content? The answer is that two of these ingredients almost cancel each other. We now demonstrate this feature.

\begin{figure}
\centering
\begin{tabular}{ccc}
\includegraphics[height=30mm,width=60mm]{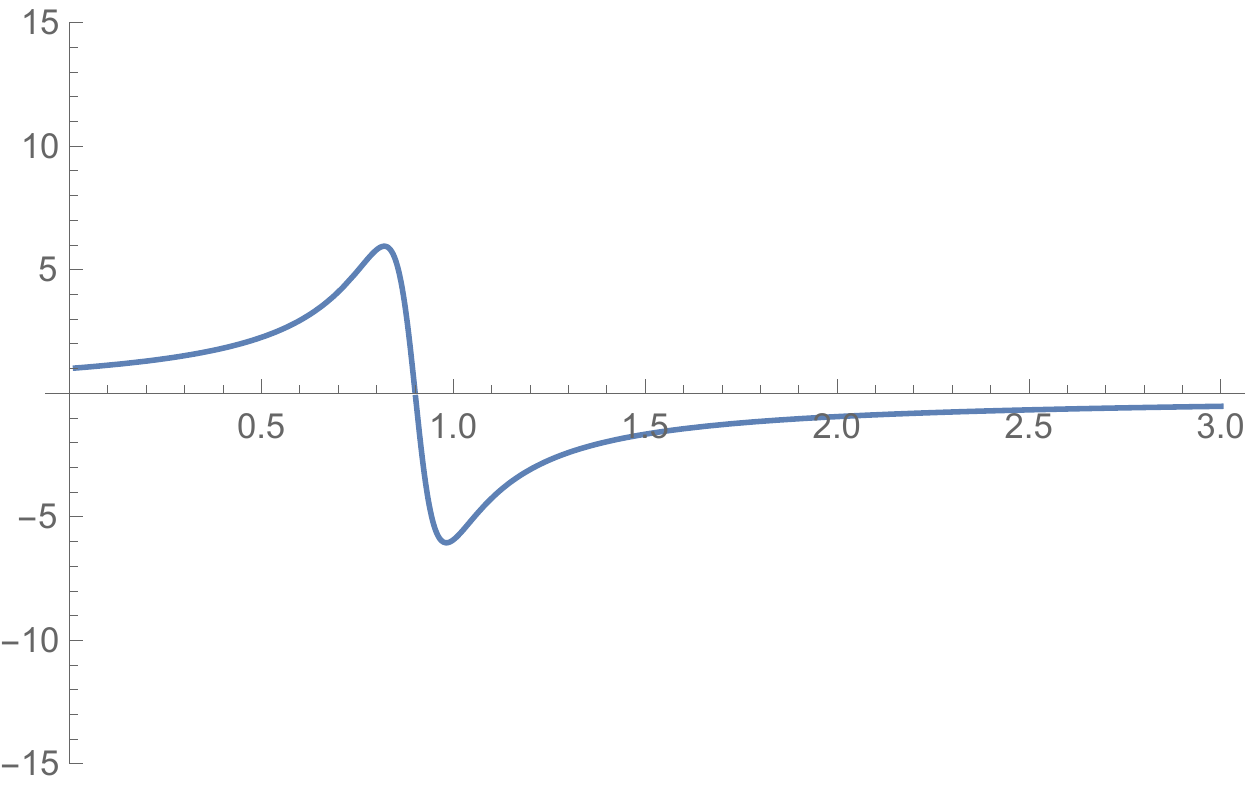}&
\includegraphics[height=30mm,width=60mm]{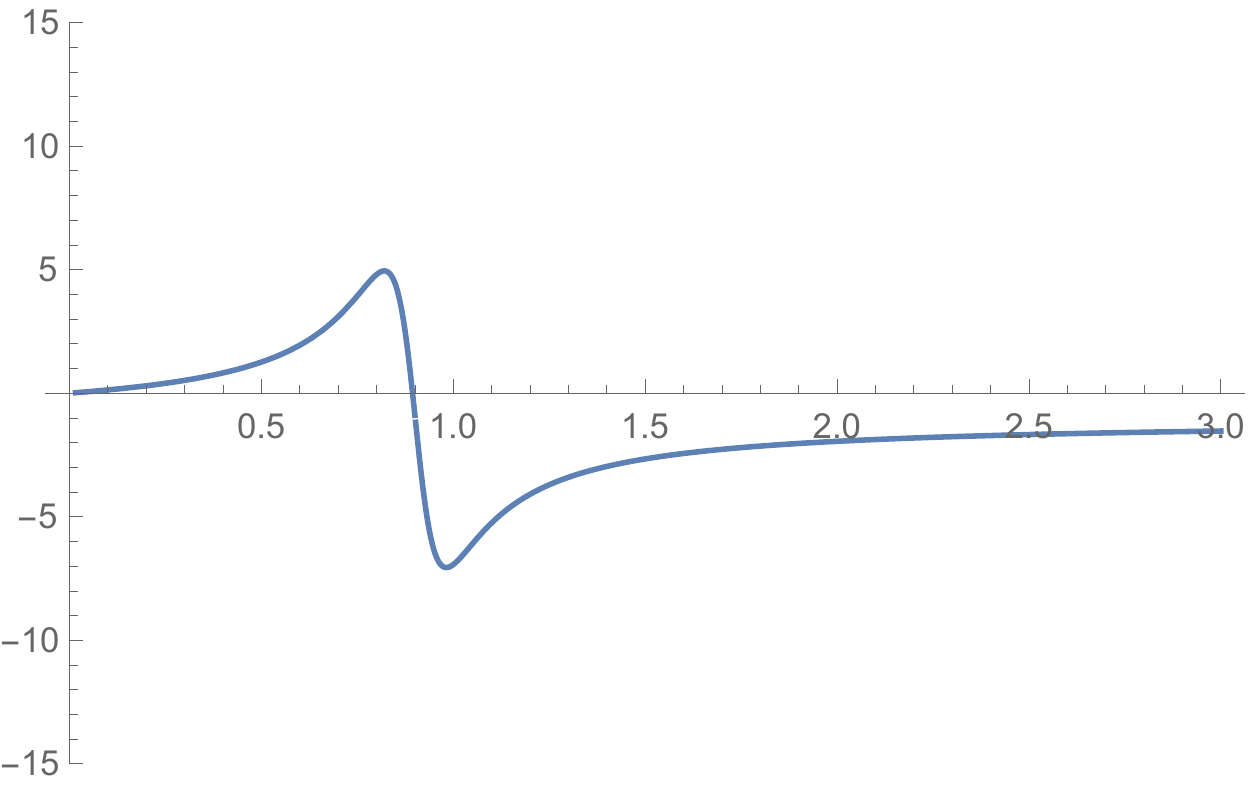}&
\includegraphics[height=30mm,width=60mm]{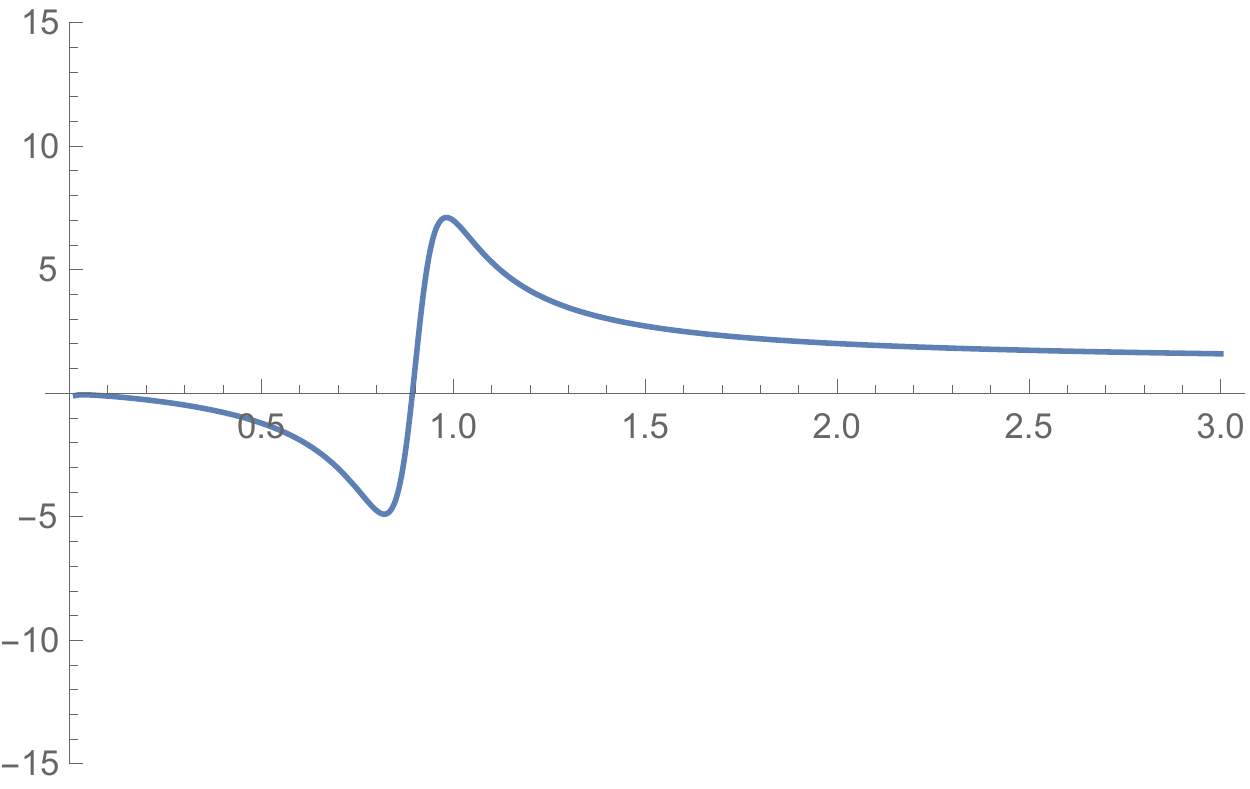}
\end{tabular}
\caption{Real parts of the three components of the propagator (left to right), the pole, the complex congugate pole and the spectral contribution. }
\label{realparts}
\end{figure}

\begin{figure}
\centering
\begin{tabular}{ccc}
\includegraphics[height=30mm,width=60mm]{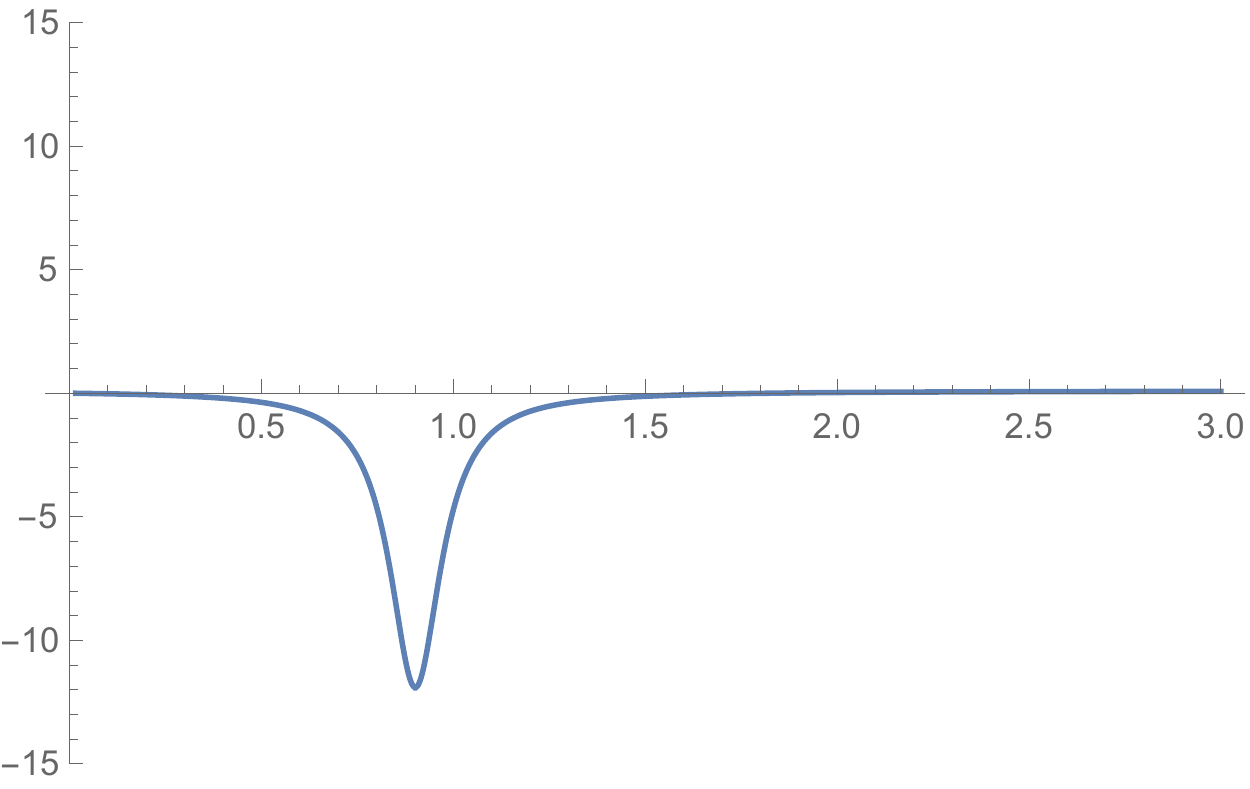}&
\includegraphics[height=30mm,width=60mm]{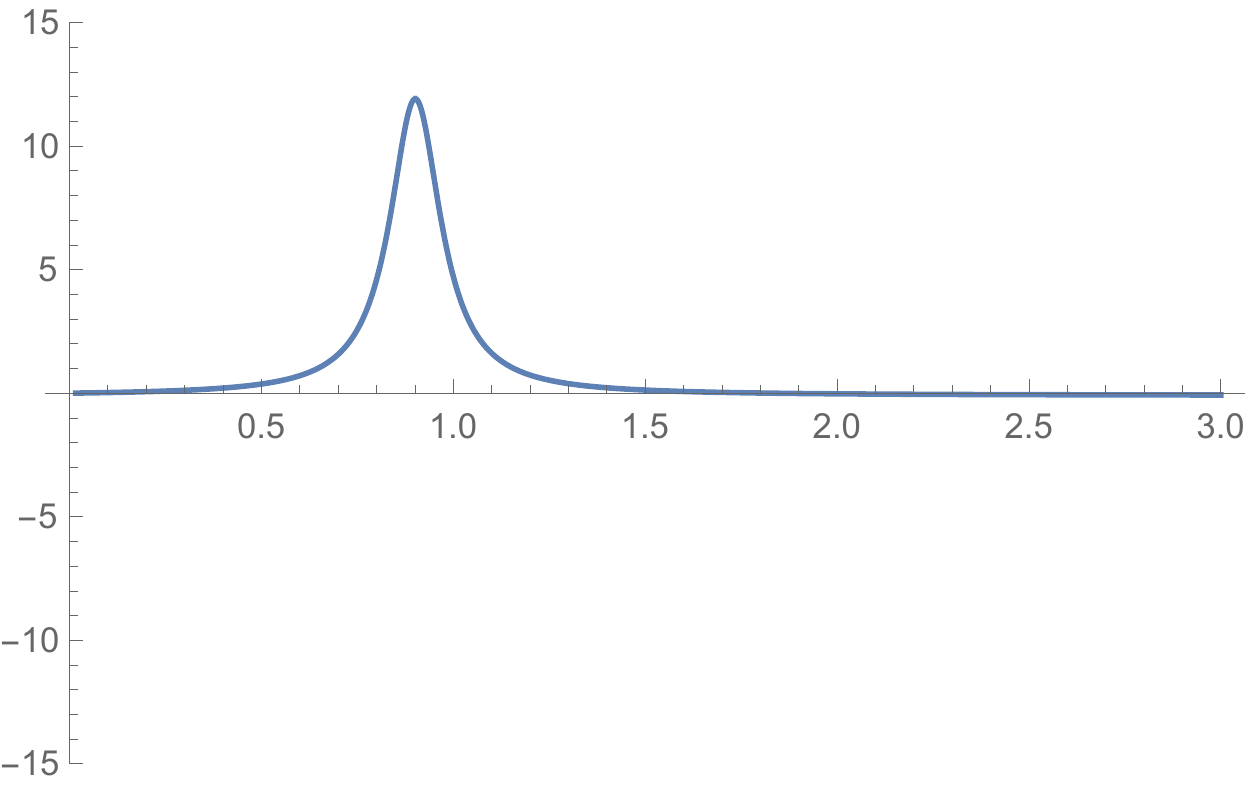}&
\includegraphics[height=30mm,width=60mm]{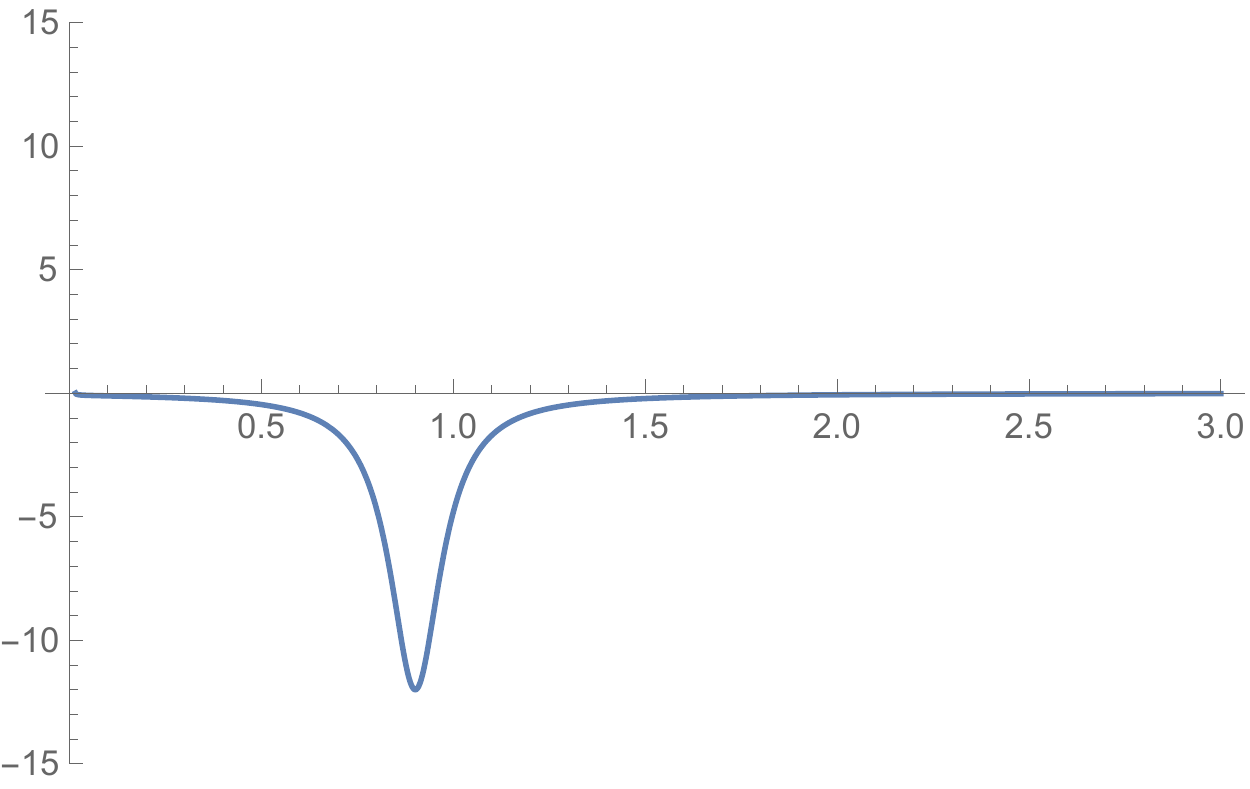}
\end{tabular}
\caption{Imaginary parts of the three components of the propagator (left to right), the pole, the complex congugate pole and the spectral contribution. }
\label{imaginaryparts}
\end{figure}

\begin{figure}[htb]
\begin{center}
\includegraphics[height=60mm,width=120mm]{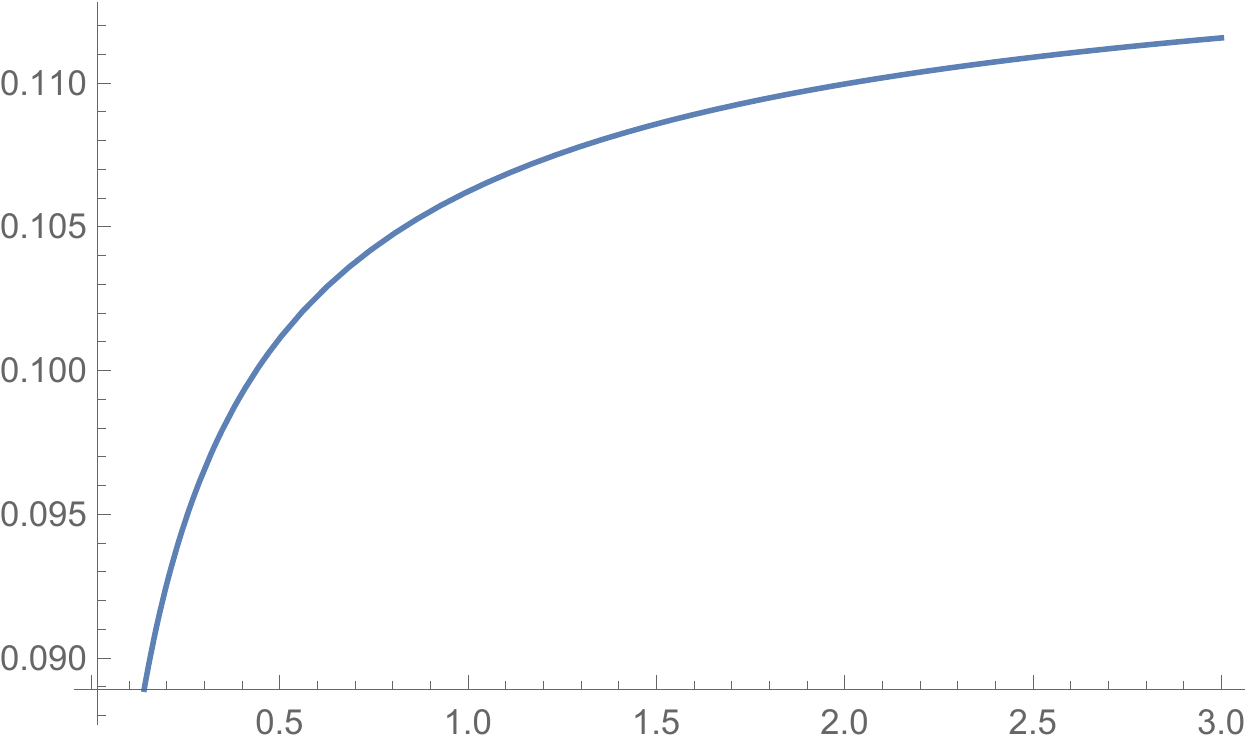}
\caption{The absolute value of the sum of the complex-conjugate pole contribution and the spectral function contribution to the propagator, multiplied by $q^2$, for $\xi^2=10$. The effect of the resonance cancels in the sum, leaving a small residual.  }
\label{diffofpoleandspectral}
\end{center}
\end{figure}

The three massive components of the propagator can be studied separately. The real parts of each are shown in Figure \ref{realparts} and the imaginary parts in Figure \ref{imaginaryparts} for our benchmark parameters. The easy comparison is that the pole and the complex-conjugate pole have the same real parts and opposite imaginary parts, as clearly required. The more interesting comparison is that the complex-conjugate pole and the spectral contribution appear to the eye to be opposite for both the real and imaginary parts. In practice, they are not exactly equal. The sum of these two components is displayed in Figure \ref{diffofpoleandspectral}. Note the change in scale on the vertical axis. There is a small residual which is required to properly describe the momentum dependence of the propagator away from the poles. But the important factor is that there is no pole-like feature in the difference. The pole structure of the complex-conjugate pole is fully removed by the contribution of the spectral function.

This important role of the spectral function is missing from the early Lee-Wick literature. As far as we know, this cancelation was first noted in Ref. \cite{Grinstein:2008bg}.

Note also that the spectral integral is important in the sum rule that determines the asymptotic behavior of the propagator. Since in Lee-Wick theories the propagator falls off as $1/q^4$ at high momentum, and each of the components in the modified Lehmann representation falls off as $1/q^2$, one must have the normalization sum rule \cite{Coleman}
\beq \label{norm}
0= 1 - \beta - \beta^* + \frac{1}{\pi} \int_{4m_f^2}^{\infty} \rho(s) ds
\eeq
Since $\beta, ~~\beta^*$ are both near unity, the spectral integral can never be neglected. For our parameters we find
\beq
\frac{1}{\pi} \int_{4m_f^2}^{\infty} \rho(s) ds  = 1.13844
\eeq
so that the normalization sum-rule is numerically satisfied.

Despite the near cancelation of two of the main ingredients of the modified Lehmann representation, the representation is important conceptually. This is largely because it demonstrates that the imaginary parts of the propagator are ``normal''. When dealing with the unstable ghost, we found that the imaginary part of the pole position was opposite of the usual expectation. Both the ghost-like nature and the imaginary part of the pole position lead to concerns about unitarity and the stability of the theory. However, in the modified Lehmann representation we see that the imaginary parts behave normally. They follow from a positive definite spectral function, with the usual $i \epsilon$ prescription in the denominator. The imaginary parts that come from the two ghost-like poles cancels each other. While there remain novel aspects of Lee-Wick theories, this feature is responsible for the indication that unitarity is still preserved in such theories.

\subsection{Analytic approximation for weak coupling}

In Lee-Wick theories the most important role of the vacuum polarization is to provide a width for the high mass ghost state, such that it is unstable. This feature arises at first order in $\alpha$. However, if one tries to solve exactly for the pole position and residue, one finds that they are non-linear in $\alpha$. Moreover, there is a large logarithm of the form $\alpha \log (\Lambda^2/m_f^2)$. For small coupling, it would be preferable to treat the propagator to the leading relevant order in the coupling, and treat further corrections in perturbation theory. Such a treatment has a few modest subtle features, such as the fact that the normalization sum-rule of Eq. \ref{norm} will only be satisfied to order $\alpha$. We describe the perturbative treatment in this section.

\begin{figure}[htb]
\begin{center}
\includegraphics[height=60mm,width=120mm]{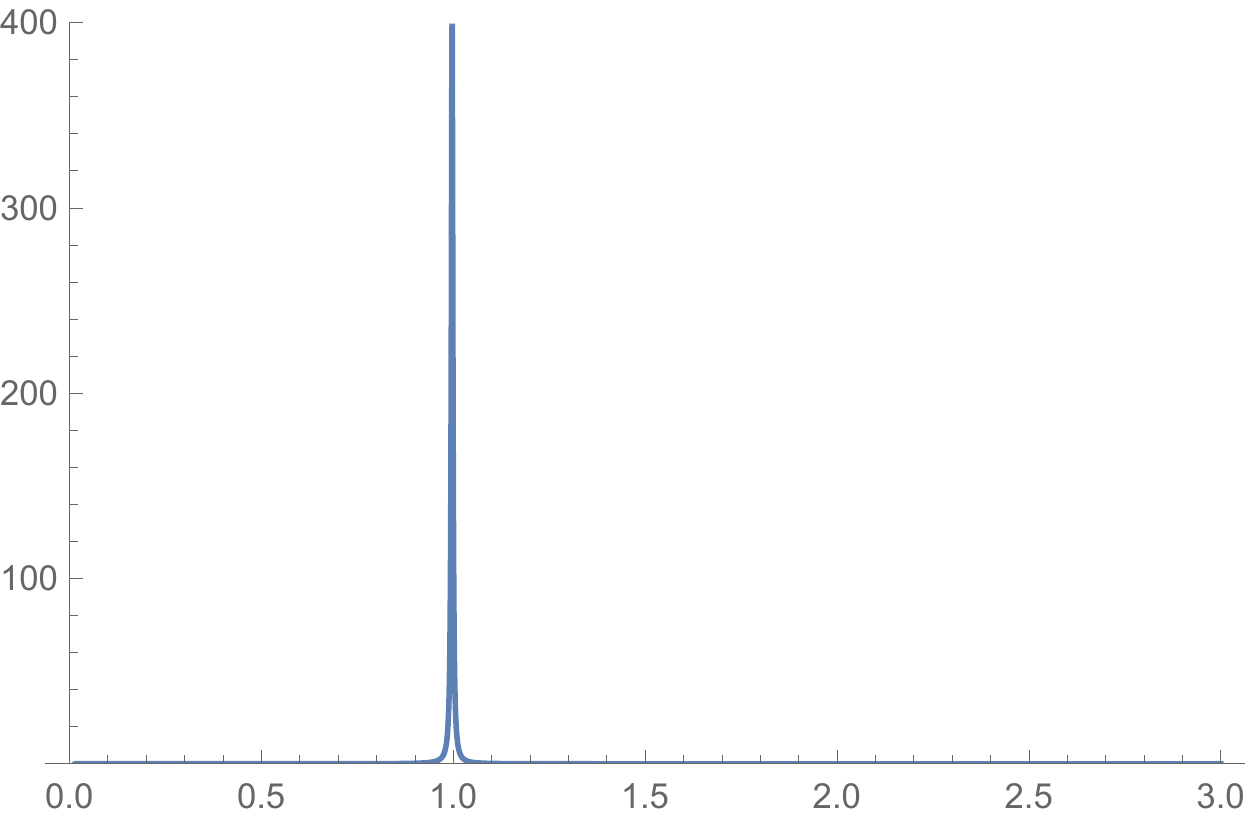}
\caption{The spectral function for $\alpha =1/137$. }
\label{spectral137}
\end{center}
\end{figure}

The origin of the large logarithm comes from the choice to define the renormalized electric charge at $q^2=0$. The residual vacuum polarization function has been defined such that $\hat{\Pi}(q^2=0) = 0$. At large values of $q^2$ it has the asymptotic expansion
\beq
\hat{\Pi}(q^2) = -\frac{\alpha}{3\pi}\left[ \log (-q^2/m_f^2) -\frac53 \right]
\eeq
Physically this corresponds to the effect of a running coupling constant. To deal with it one can include the coupling constant in the propagator. More explicitly, at large positive $q^2$
\beq
\alpha D(q^2) \sim \frac{\alpha}{q^2 \left[1 -\frac{\alpha}{3\pi}[ \log (q^2/m_f^2) -\frac53 ] - \frac{q^2}{\Lambda^2}  + i \frac{\alpha}{3}\right] }
\eeq
Written in terms of the running coupling defined at the value $q^2=\Lambda^2$, we have instead
\beq
\alpha D(q^2) \sim \frac{\alpha (\Lambda)}{q^2 \left[1 -\frac{\alpha}{3\pi} \log (q^2/\Lambda^2)  - \frac{q^2}{\Lambda^2}  + i \frac{\alpha}{3}\right] }
\eeq
to first order in the coupling. The large logarithm has been absorbed into the coupling and the remaining logarithmic running is weak near the pole.

Locating the pole is now easy to first order in the coupling. The real part of the propagator vanishes at $q^2=\Lambda^2$ for real $q^2$.
It remains at this value to first order in $\alpha$ even when $q^2$ is allowed to be complex. The resonance parameters at leading order are
\begin{eqnarray}\label{weak}
M^2 &=& m_p^2+i \gamma  \nonumber \\
m_p^2 &=& \Lambda^2  \nonumber \\
{\gamma} &=& \frac{\alpha}{3} \nonumber \\
{\beta} &=& 1\ \ .
\end{eqnarray}

At weak coupling, the poles and the spectral function are very narrow. For the physical value of $\alpha$, but other parameters unchanged from the previous section, the spectral function is shown in Fig. \ref{spectral137}. Nevertheless, when it is integrated over all values of $s$, the result differs from unity by an amount of order $\alpha$. In this case, the normalization sum rule is not satisfied exactly. However, given that the spectral function is so highly peaked and narrow, one can easily use a normalized Breit-Wigner distribution as an approximation, which then satisfies the sum rule.

The modified Lehmann representation continues to have three massive pole-like structures: the physical pole, its complex conjugate and the Breit-Wigner of the spectral function integral. In the narrow width limit, the spectral integral can be done explicitly. Not surprisingly, it describes a massive pole
\beq
\frac{1}{\pi}\int_{4m_f^2}^\infty ds \frac{\rho(s)}{q^2-s+i\epsilon} \sim \frac{1}{q^2- m_p^2 +i\gamma}
\eeq
and the cancellation with the complex conjugate pole becomes exact.

\section{Extending the Lee-Wick sectors}

Our starting point, Eq. \ref{basicform} from \cite{Boulware:1983vw}, captures the key feature of the Lee-Wick theory. However, in general such theories are more complicated. For example, Lee-Wick QED also contains a heavy fermion ghost which appears in the fermion propagator \cite{Lee:1970iw}. If that ghost were light enough, it could modify the imaginary part of the vacuum polarization function at energies below the photon ghost pole. So our starting point corresponds to the case where the $\Lambda$ parameter for the heavy fermion ghost is larger than that in the photon propagator. Loops of the heavy fermion would renormalize the parameters of the low energy limit but would not generate an imaginary piece in the propagator at these energies, and our analysis of the pole structure would be unchanged.

To leading order, the analysis of poles in the fermion propagator would share the same features as the photon analysis above. The heavy fermion ghost picks up an imaginary part due to the coupling to the light states in the theory. There would be a modified Lehmann representation for the fermion propagator also, with a similar result.

If the heavy fermions were lower in mass than the photon ghost, unitarity requires that they generate imaginary parts also in the vacuum polarization. This would modify the location of the pole and the width, but would leave the rest of the analysis unchanged. It is expected that Lee-Wick theories are unitary. The clearest calculation that we know of demonstrating unitarity and showing the role of imaginary parts from heavy ghosts is in Ref. \cite{Grinstein:2008bg} within $O(N)$ theories.

\section{Summary}

Most discussions of Lee-Wick theories emphasize the existence of an unstable massive ghost pole and its complex conjugate. The point of this paper is to emphasize that the spectral function also has a pole-like structure - it is close to a Breit Wigner shape. In addition, the effect of the spectral integral is to cancel the resonance behavior of the complex-conjugate pole, leaving in general a small (non-resonant) residual.

Nevertheless, the modified Lehmann representation is conceptually and calculationally useful. The imaginary parts of the propagator at high energy lie in the spectral integral, which has the usual $i\epsilon$ structure. It is important for real calculations to include the spectral integral when discussing loops. We do not address here anything about the unusual contours chosen when evaluating Feynman integrals in these theories. However, the Lehmann representation captures the physics of Lee-Wick theories when one includes the resonance structure in the spectral function.

Much of the present interest in Lee-Wick theories comes from the study of quantum gravity \cite{Stelle:1976gc, Julve:1978xn, Fradkin:1981hx, Tomboulis, Smilga, Einhorn, Strumia, Donoghue, Menezes, Holdom, Mannheim, Hooft, Shapiro, Narain, Anselmi, Alvarez-Gaume, Modesto, Accioly, Salvio}. If it is possible to describe a UV complete theory of quantum gravity using renormalizable quantum field theory, the gravitational interaction will necessarily involve higher derivatives
in the fundamental action. This in general leads to high mass ghost-like states. Analogous to the discussion above, these states will be unstable and will decay into the light particles of the theory. The propagators of the gravity theory will be similar to that described above, with some differences due to the momentum dependence in the gravitational vacuum polarization. In particular, the compensation found in the modified Lehmann representation and the need for a resonance structure in the spectral integral will be the same. The understanding of the propagator is an important first step in the exciting possibility that quantum gravity can be described by a renormalizable quantum field theory.

\section*{Acknowledgements} We are grateful for correspondence on these topics with Ben Grinstein, Donal O'Connell, Mark Wise and Terry Tomboulis. This work has been supported in part by the National Science Foundation under grant NSF PHY15-20292 and PHY1820675 (JFD), Conselho Nacional de Desenvolvimento Cient\'ifico e Tecnol\'ogico - CNPq under grant 307578/2015-1 (GM) and Funda\c{c}\~ao Carlos Chagas Filho de Amparo \`a Pesquisa do Estado do Rio de Janeiro - FAPERJ under grant E-26/202.725/2018 (GM).

\end{document}